# Superconductivity in In-doped AgSnBiTe$_3$ with possible band inversion

Tsubasa Mitobe[1], Kazuhisa Hoshi[1], Md. Riad Kasem[1], Ryosuke Kiyama[1], Hidetomo Usui[2], Aichi Yamashita[1], Ryuji Higashinaka[1], Tatsuma D. Matsuda[1], Yuji Aoki[1], Takayoshi Katase[3], Yosuke Goto[1], Yoshikazu Mizuguchi[1]*

[1]Department of Physics, Tokyo Metropolitan University, 1-1, Minami-osawa, Hachioji 192-0397
[2]Department of Physics and Materials Science, Shimane University, 1060, Nishikawatsucho, Matsue 690-8504, Japan.
[3]Laboratory for Materials and Structures, Institute of Innovative Research, Tokyo Institute of Technology, 4259 Nagatsuta, Midori, Yokohama 226-8503, Japan

Corresponding author: Yoshikazu Mizuguchi (mizugu@tmu.ac.jp)

## Abstract

We investigated the chemical pressure effects on structural and electronic properties of SnTe-based material using partial substitution of Sn by Ag$_{0.5}$Bi$_{0.5}$, which results in lattice shrinkage. For Sn$_{1-2x}$(AgBi)$_x$Te, single-phase polycrystalline samples were obtained with a wide range of $x$. On the basis of band calculations, we confirmed that the Sn$_{1-2x}$(AgBi)$_x$Te system is basically possessing band inversion and topologically preserved electronic states. To explore new superconducting phases related to the topological electronic states, we investigated the In-doping effects on structural and superconducting properties for $x$ = 0.33 (AgSnBiTe$_3$). For (AgSnBi)$_{(1-y)/3}$In$_y$Te, single-phase polycrystalline samples were obtained for $y$ = 0–0.5 by high-pressure synthesis. Superconductivity was observed for $y$ = 0.2–0.5. For $y$ = 0.4, the transition temperature estimated from zero-resistivity state was 2.4 K, and the specific heat investigation confirmed the emergence of bulk superconductivity. Because the presence of band inversion was theoretically predicted, and the parameters obtained from specific heat analyses were comparable to In-doped SnTe, we expect that the (AgSnBi)$_{(1-y)/3}$In$_y$Te and other (Ag,In,Sn,Bi)Te phases are candidate systems for studying topological superconductivity.

# Introduction

Metal tellurides (*M*Te) with a NaCl-type structure have been extensively studied due to their physical properties as topological materials[1–4], thermoelectric materials[5–7], and superconductors[8–12]. Among them, In-doped SnTe superconductors have been drawing attention as a candidate system of a topological superconductor[4,13–17]. SnTe is a topological crystalline insulator, and superconductivity is typically induced by In doping at the Sn site. The superconducting transition temperature ($T_c$) of (Sn,In)Te increases by In doping. Although a simple picture proposes that doped In acts as a hole dopant, detailed analyses of carrier characteristics, superconducting properties, and electronic states of (Sn,In)Te revealed that the In doping does not simply act as a dopant of holes, but the superconductivity emerges in a regime where electron carriers are dominant[16]. Therefore, to understand the nature and the mechanisms of superconductivity in SnTe-based, development of new superconductors based on NaCl-type tellurides is important.

In *M*Te, the *M* site can be alloyed flexibly. For example, single crystals (films) of (Sn,Pb)Te can by grown with a wide solution range, and the alloy system has provided a platform to study topological nature of *M*Te[1,18,19]. Because the Pb substitution for SnTe expands the lattice, which corresponds to negative chemical pressure at the *M*-Te bond, contrasting positive chemical pressure in *M*Te leads the way for further expanding the research field of superconductivity in *M*Te. In addition, although In-doped systems of (In,Sn,Pb)Te have been studied as a topological superconductor candidate[20], there has been no detailed study on superconducting properties and crystal structure of NaCl-type MTe with a lattice constant smaller than SnTe. In this study, we focused on the (Ag,Sn,Bi)Te system that has been studied as a thermoelectric material and found that the (Ag,Sn,Bi)Te system is possible topological material[21,22]. In SnTe, Sn is divalent, $Sn^{2+}$. When $Ag^{+}_{0.5}Bi^{3+}_{0.5}$ substitutes $Sn^{2+}$, the total valence states (charge neutrality) has been preserved. Therefore, the $Ag_{0.5}Bi_{0.5}$ substitution for the Sn site is successfully achieved in a wide range up to the end member of $AgBiTe_2$. Here, we show the evolution of the structural and electronic characteristics of $Sn_{1-2x}(AgBi)_xTe$. Then, we report superconductivity induced by In substitution in $(AgSnBi)_{(1-y)/3}In_yTe$.

# Results and discussion
## Structural and electronic characteristics of $Sn_{1-2x}(AgBi)_xTe$

As mentioned above, the (Ag,Sn,Bi)Te system would be an important system to expand the material variety of SnTe-based compounds including new superconductors. We started this work by investigating lattice compression in $Sn_{1-2x}(AgBi)_xTe$, in which $x$ corresponds to the total amount of Ag and Bi substituted for the Sn site in the SnTe structure. Polycrystalline samples of Sn$_{1-2x}$(AgBi)$_x$Te were synthesised by a melting method. Figure 1a shows the X-ray diffraction (XRD) patterns for Sn$_{1-2x}$(AgBi)$_x$Te. The XRD peaks correspond to that expected for the NaCl-type structure (Fig. 1b) and systematically shift to higher angles, which indicates lattice shrinkage with increasing $x$. The lattice constant $a$ was determined by Rietveld refinements and plotted in Fig. 1c. The trend is consistent with previous reports[20,21]. In the refinements, Ag, Sn, and Bi were assigned to the $M$ site as shown in Fig. 1b, and the nominal composition (fixed) was used. Since we used laboratory XRD in this study, isotropic displacement parameter $B_{iso}$ was fixed to 1 for all sites.

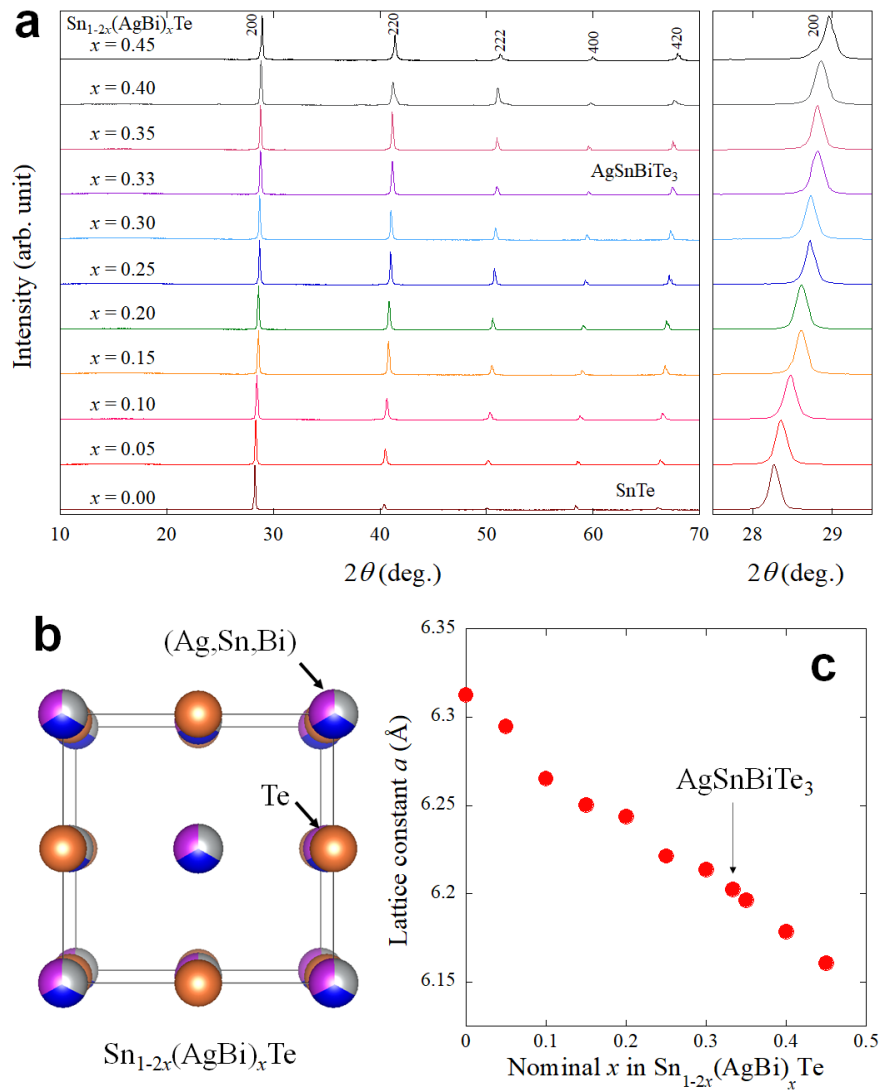

**Fig. 1. Crystal-structure evaluation of Sn$_{1-2x}$(AgBi)$_x$Te. a.** XRD patterns of Sn$_{1-2x}$(AgBi)$_x$Te. **b.** Schematic image of the NaCl-type crystal structure of Sn$_{1-2x}$(AgBi)$_x$Te. **c.** $x$ dependence of the lattice constant.

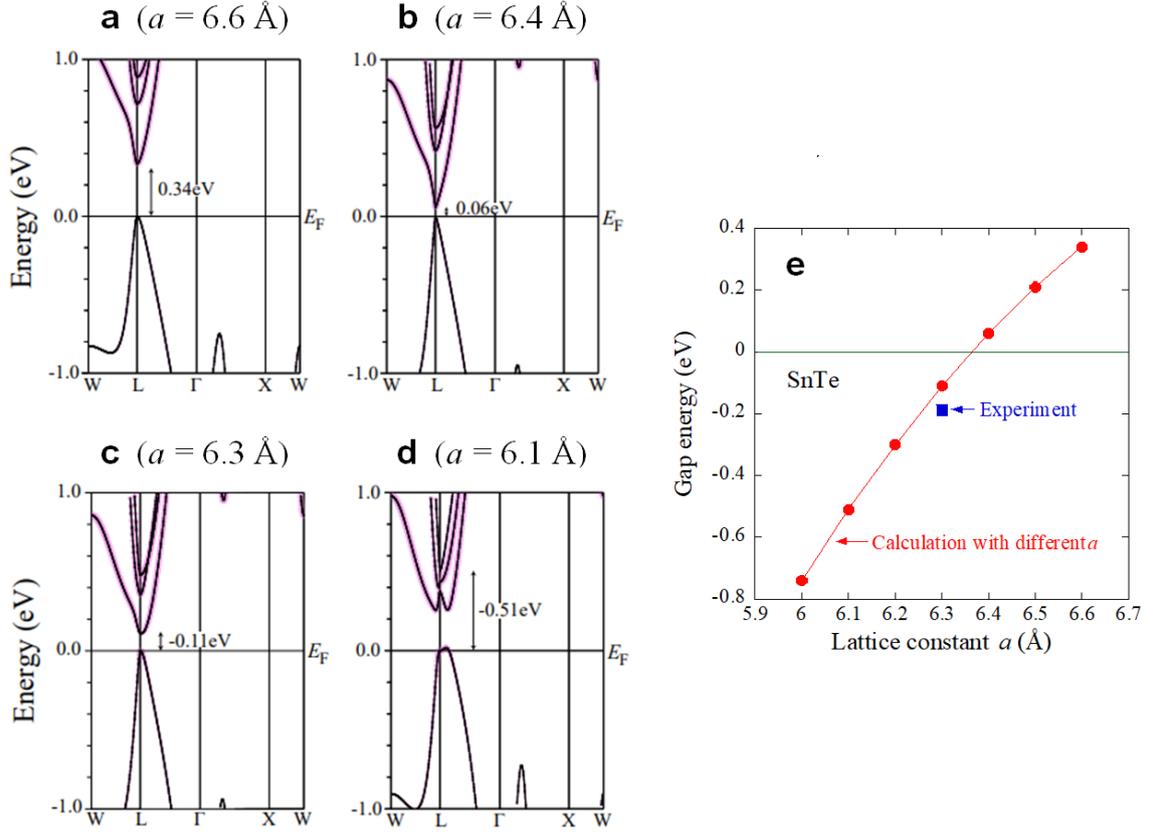

**Fig. 2. Electronic structure for SnTe with various lattice constants. (a-d)** Calculated band structure. Contribution of Sn-*p* orbitals is represented by the size of pink-circle symbol. According to the calculations, a band inversion occurs with lattice constants smaller than 6.35 Å as plotted in Fig. **2e**. Experimental data in Ref. 23 is plotted. Gap energy was calculated from the lowest energy of the Sn-p orbital and the highest energy of the Te-*p* orbital at the L point.

Figure 2 shows the calculated band structure for SnTe with various lattice constants of 6.6, 6.4, 6.3 (close to the lattice constant of SnTe), and 6.1 Å. By calculating the lowest energy of the Sn-*p* orbital and the highest energy of the Te-*p* orbital at the L point in reciprocal space, we confirmed that a band inversion transition occurs at around $a$ = 6.35 Å (Fig. 2e). Note that the contributions of Sn-*p* orbitals are represented by the size of the circle symbols in Figs. 2a–d. Furthermore, calculation results with a smaller lattice constant show that SnTe-based materials with a small lattice constant of 6.1 Å also shows a band inversion. The observed trend is consistent with a previous theoretical work on SnTe and PbTe systems[2]. Therefore, we consider that the current system of $Sn_{1-2x}(AgBi)_x Te$ basically possesses a topologically-preserved band structure with a wide range of *x*. Although calculations in this work have been performed on SnTe with different lattice constants, no obvious modification of the band structure is expected when Sn was partially replaced by $Ag_{0.5}Bi_{0.5}$ in real materials because the difference in spin-orbit interactions expected from those elements is not large. Since the orbital characteristics in the band structure does not largely change between SnTe and $Sn_{1-2x}(AgBi)_xTe$, we consider that the topological invariant in this system is mirror Chern number from the analogy to SnTe, which suggests that the $Sn_{1-2x}(AgBi)_xTe$ system is a potential topological crystalline insulator. On the basis of the investigations of lattice constant and band structure for the SnTe-based system $Sn_{1-2x}(AgBi)_xTe$, we selected $AgSnBiTe_3$ ($x = 0.33$) for a parent phase in which In-substitution effects are examined in this study.

## Superconducting properties of $(AgSnBi)_{(1-y)/3}In_yTe$

For the In-doped $AgSnBiTe_3$ system (see Fig. 3c for crystal structure), we used a chemical formula, $(AgSnBi)_{(1-y)/3}In_yTe$, because the In amount doped to the parent phase of $AgSnBiTe_3$ can be easily understood. Figure 3a shows the powder XRD patterns of $(AgSnBi)_{(1-y)/3}In_yTe$; the In-doped samples were synthesised by high-pressure annealing. As shown in Fig. 3b, Rietveld refinement reveals that tiny (shoulder) anomaly was observed. Such a shoulder structure would be due to the presence of inhomogeneous regime with a slightly different lattice constant and was observed in NaCl-type tellurides containing multiple $M$-site elements[24,25]. In particular, we tested several annealing conditions for $y = 0.4$, and the condition described in the **Method** section was found to be the best. As shown in Fig. 3d, the lattice constant decreases with increasing $y$, which is a trend similar to other In-doped M-Te systems[8,12]. The actual element concentrations of the samples were examined by energy-dispersive X-ray spectroscopy (EDX), and the results are shown in Fig. S1 (supplementary information). Basically, the actual compositions were close to the nominal values.

In Fig. 4, the temperature dependence of electrical resistivity of $AgSnBiTe_3$ ($x = 0.33$, $y = 0$) is displayed. Resistivity slightly decreases with decreasing temperature, and an increase in resistivity was observed at low temperatures. The result is consistent with the band calculations, based on SnTe lattice, in Fig. 2, and hence, the parent phase $AgSnBiTe_3$ would possess a narrow band gap. Since the calculated gap energy (Fig. 2e) indicates that a band inversion is expected for a metal telluride with $a = 6.2$ Å, $AgSnBiTe_3$ with $a = 6.20217(9)$ Å is expected to have topologically preserved electronic states near the Fermi energy ($E_F$). Thus, the In-substitution effects on physical properties of $AgSnBiTe_3$ are of interest because of the analogy to (Sn,In)Te, in which topological superconductivity is expected to emerge. As expected, In-doped $AgSnBiTe_3$ shows superconductivity as displayed in Fig. 4. The resistivity data for $y = 0.4$ shows a metallic behavior and zero resistivity was observed at $T_c^{zero} = 2.4$ K. For the other samples with different In concentration, the temperature dependence of resistivity are shown in Fig. S2 (Supplemental Information). The sample with $y = 0.1$ shows almost no temperature dependence in resistivity, but other samples ($y$ = 0.2–0.5) show metallic conductivity.

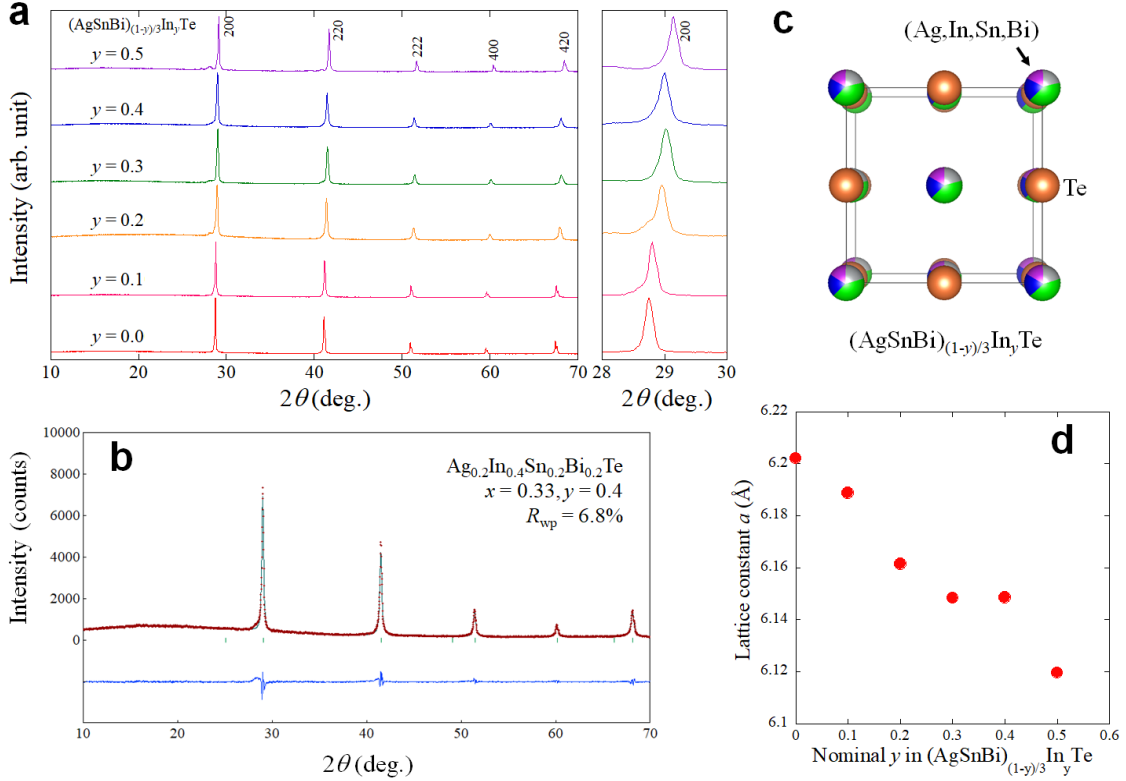

**Fig. 3. Crystal-structure evaluation of (AgSnBi)$_{(1-y)/3}$In$_y$Te. a.** XRD patterns of (AgSnBi)$_{(1-y)/3}$In$_y$Te. **b.** Rietveld refinement result for $y$ = 0.4. **c.** Schematic image of the crystal structure. **d.** $y$ dependence of the lattice constant.

To investigate upper critical field ($B_{c2}$), temperature dependences of resistivity were measured under various magnetic fields up to 1.5 T as plotted in Fig. 5a. For the estimation of $B_{c2}$, the transition temperature was determined as a temperature where resistivity drops to 50% of the normal-state resistivity. Using the WHH model (Werthamer-Helfand-Hohenberg model)[26], which is applicable for a dirty-limit type-II superconductor, the $B_{c2}(0)$ was estimated as 1.2 T, as shown in Fig. 5b.

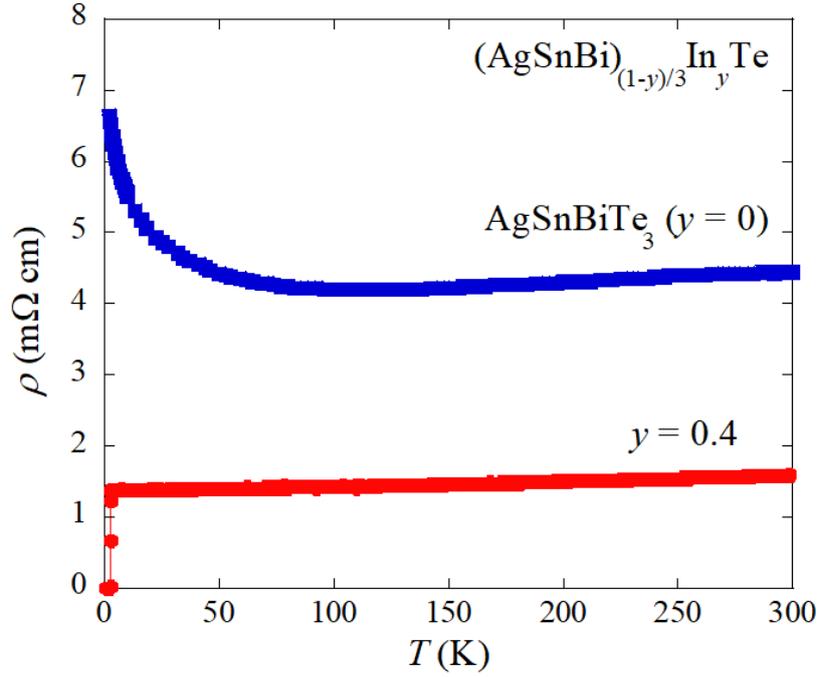

**Fig. 4. Temperature dependence of electrical resistivity for $y = 0$ (AgSnBiTe$_3$) and $y = 0.4$.** In-doped ($y = 0.4$) sample shows a superconducting transition at $T = 2.4$ K.

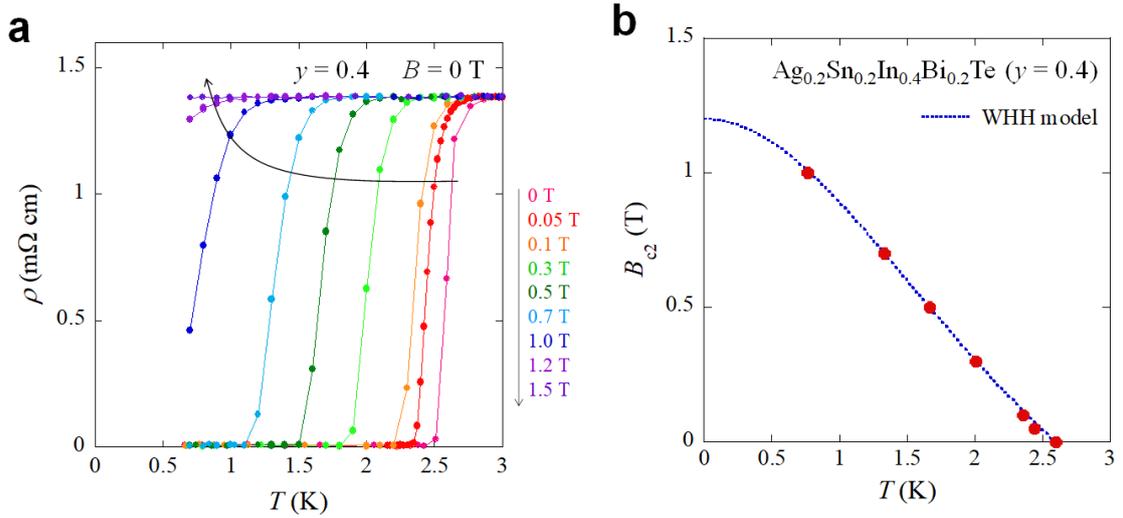

**Fig. 5. Estimation of upper critical field $B_{c2}$ for $y = 0.4$. a.** Temperature dependences of resistivity under magnetic fields up to 1.5 T. With increasing field, the $T_c$ systematically decreases. **b.** Temperature dependence of $B_{c2}$. For the plot of $B_{c2}$, middle-point temperatures, where resistivity became 50% of normal-state resistivity at 3 K, were estimated in Fig. 5a. The dotted line shows WHH fitting.

Figure 6 shows the analysis results of specific heat measurements. In Fig. 6a, specific heat data under 0 and 5 T in a form of $C/T$ are plotted as a function of $T^2$. From the data under 5 T, the electronic specific heat parameter ($\gamma$) and the Debye temperature ($\theta_D$) were estimated as 2.35(5) mJmol$^{-1}$K$^{-2}$ and 186 K, respectively. The low-temperature specific-heat formula with anharmonic term was used for the analysis: $C = \gamma T + \beta T^3 + \delta T^5$, where $\beta T^3$ is the lattice contribution to the specific heat, and the $\delta T^5$ term accounts for anharmonicity of the lattice. The $\theta_D$ was calculated from $\beta = (12/5)\pi^4(2N)k_B\theta_D^{-3}$, where $N$ and $k_B$ are the Avogadro constant and Boltzmann constant, respectively. To characterize the superconducting properties, the electronic contribution ($C_{el}$) under 0 T, which was calculated by subtracting lattice contributions from the specific heat data under 0 T, is plotted in the form of $C_{el}/T$ as a function of temperature in Fig. 6b. The clear jump of $C_{el}/T$ and decrease in $C_{el}/T$ at low temperatures suggest the emergence of bulk superconductivity. The superconducting jump in $C_{el}$ ($\Delta C_{el}$) estimated with $T_c$ = 2.16 K is 1.34$\gamma T_c$, which is comparable to the value expected from a full-gap superconductivity based on the BCS model ($\Delta C_{el} = 1.43\gamma T_c$)[27]. Those values obtained from specific heat are similar to those obtained for In-doped SnTe and Ag-doped SnTe superconductors with 20% In (or Ag) doping[8,28].

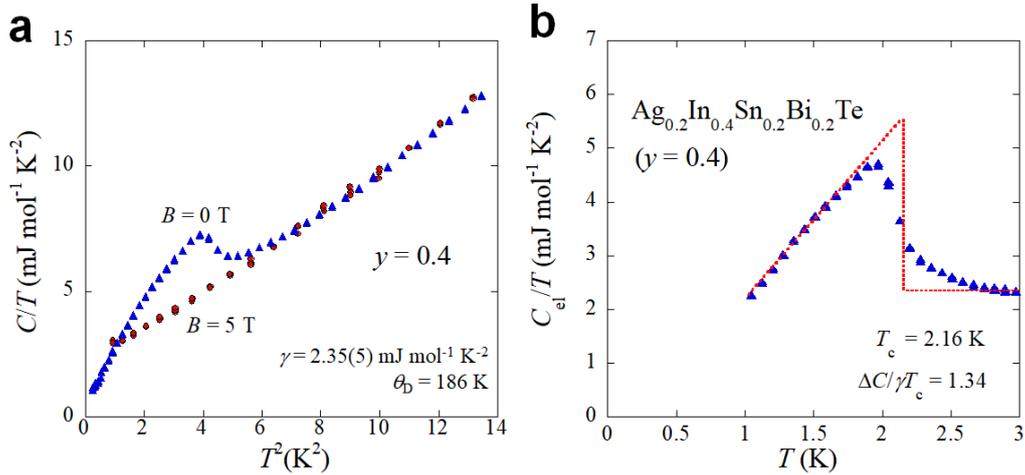

**Fig. 6. Specific heat data for $y = 0.4$. a.** Specific heat ($C/T$) data under magnetic fields ($B$) of 0 and 5 T plotted as a function of $T^2$. By fitting the data under $B$ = 5 T, electronic specific heat parameter ($\gamma$) and Debye temperature ($\theta_D$) were estimated. **b.** Temperature dependence of the electronic contribution ($C_{el}/T$) under $B$ = 0 T. The electronic contribution was calculated by subtracting the lattice contribution. $T_c$ and the superconducting jump at $T_c$ ($\Delta C$) were estimated by considering the entropy balance, as guided by the red dashed lines. $\Delta C/\gamma T_c$ = 1.34, which is close to BCS value of 1.43

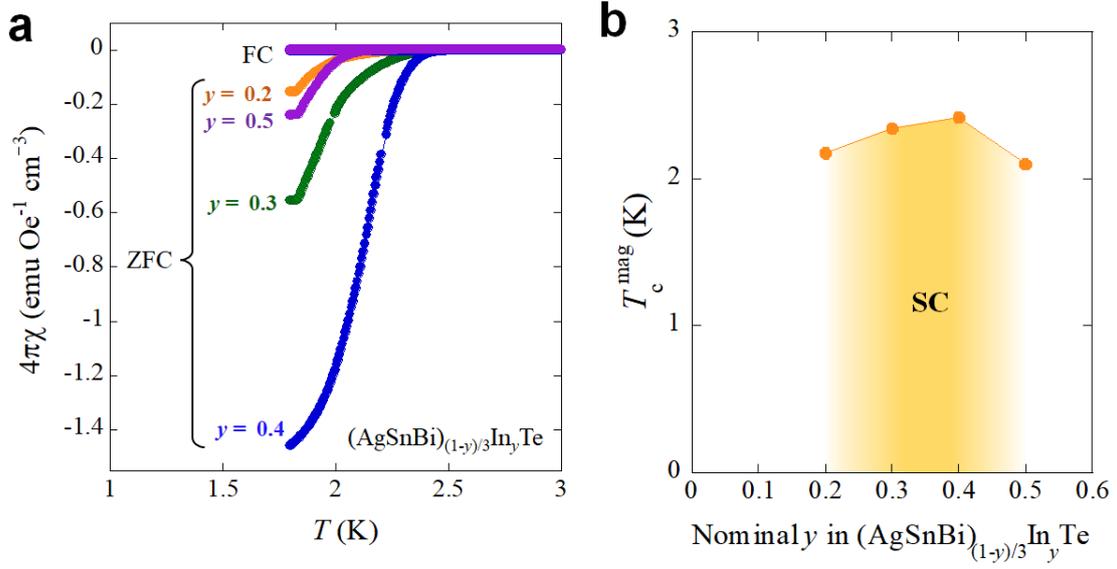

**Fig. 7. Magnetic susceptibility and superconducting transition temperature for $(AgSnBi)_{(1-y)/3}In_yTe$. a.** Temperature dependences of magnetic susceptibility ($4\pi\chi$) for $y$ = 0.2, 0.3, 0.4, 0.5. **b.** In concentration ($y$) dependence of $T_c$ estimated from 1% shielding volume fraction ($T_c^{mag}$).

To investigate the In-doping dependence of superconducting states, temperature dependences of magnetic susceptibility ($4\pi\chi$) was measured for $(AgSnBi)_{(1-y)/3}In_yTe$. For, $y$ = 0 and 0.1, no superconducting transition was observed at temperatures above 1.8 K. Superconducting diamagnetic signals were observed for $y$ = 0.2–0.5, as displayed in Fig. 7a. Particularly, samples with $y$ = 0.3 and 0.4 showed a large shielding fraction. Note that the data has not been corrected by demagnetisation effect. The observation of the largest shielding volume fraction is consistent with the bulk superconductivity confirmed in specific heat measurements. As shown in Fig. 7b, $T_c$ estimated from magnetic susceptibility ($T_c^{mag}$) becomes the highest at $y$ = 0.4 and decreases with further In doping ($y$ = 0.5).

The estimated $T_c$ for $y$ = 0.4 is lower than that for (Sn,In)Te ($T_c \sim$ 4.5 K)[16]. The difference may be caused by three possible reasons: (i) carrier concentration, (ii) the effect of disorder, and (iii) lattice constant. On (i) carrier concentration, we performed measurements of Seebeck coefficient (at room temperature) and Hall coefficient (at 5 K for $y$ = 0.4). Figure 8 shows the $y$ dependence of Seebeck coefficient ($S$). For $y$ = 0, a large positive value of $S$ was observed. This is consistent with the band calculation, indicating that the parent phase of $y$ = 0 is a semiconductor. With increasing $y$, $S$ becomes negative, and the absolute value for $y$ > 0.1 becomes less than 10 μV/K, which is a typical value of metals. In Fig. S3 (Supplemental Information), the magnetic field dependence of Hall resistance at 5 K is plotted. By linear fitting of the data and assuming a single-band model, carrier concentration was calculated as $7.4\times10^{21}$ cm$^{-3}$. These results on the evolution of carrier concentration by In substitution in $(AgSnBi)_{(1-y)/3}In_yTe$ would suggest that electrons are doped by $In^{3+}$ substitution, and the doping situation in the present system is clearly different from that observed in $(Sn,In)Te^{16}$. Therefore, to investigate the relationship between superconducting properties and carrier concentration in $(AgSnBi)_{(1-y)/3}In_yTe$, further investigation with various probes is needed. On the effect of (ii) disorder, it is a fact that high configurational entropy of mixing is present at the $M$ site as described in the following discussion. However, comparison of $T_c$ between $y = 0.4$ ($M$ = Ag,In,Sn,Bi) and other $M$Te superconductors, for example, $M$ = In ($T_c \sim 3$ K)[9,12] and $M$ = $Sn_{0.8}Ag_{0.2}$ ($T_c = 2.3$ K)[11], suggests that the higher disorder in $y = 0.4$ ($M$ = Ag,In,Sn,Bi) is not highly affecting $T_c$. On the effect of (iii) lattice constant, the electronic structure of SnTe-based materials is modified by lattice constant as shown above. In addition, in Ref. 25, we showed that $T_c$ of $M$Te shows a positive relation to lattice constant. Therefore, the $T_c$ obtained for $y = 0.4$ would be reasonable to the trend of $T_c$-lattice constant for $M$Te. According to those facts, we consider that the difference in $T_c$ between $y = 0.4$ and $(Sn,In)Te$ is caused by the difference in the electronic states (carrier concentration and/or lattice constant).

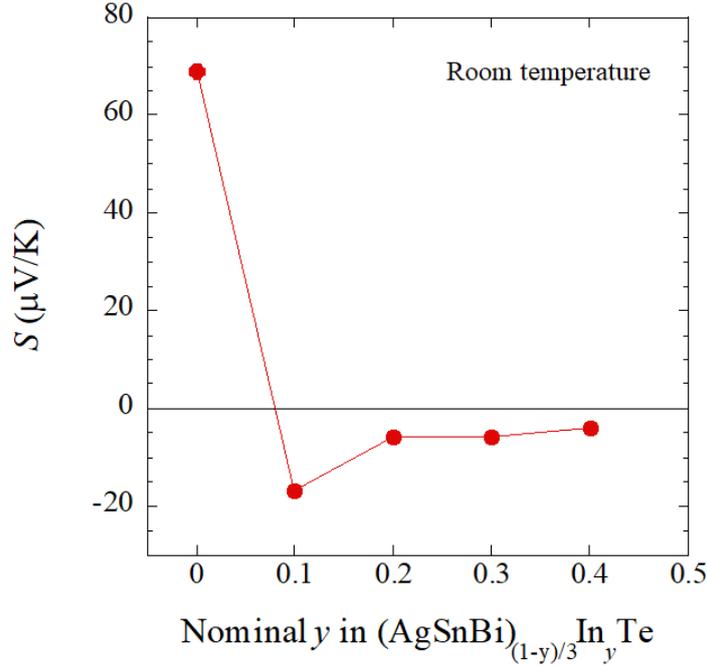

**Fig. 8. Effect of In doping on Seebeck coefficient ($S$) for $(AgSnBi)_{(1-y)/3}In_yTe$. In concentration ($y$) dependence of S at room temperature is shown.**

**Solubility limit and phase stabilisation by configurational entropy of mixing**

The doping phase diagram for superconductivity of $(AgSnBi)_{(1-y)/3}In_yTe$ is compared with that for the (Sn,In)Te and (Pb,In)Te systems in the following discussion. Having looked at the phase diagrams of (Sn,In)Te and (Pb,In)Te[8,12], superconductivity is observed in a wide range of In concentration. As shown in Fig. 7b, however, the superconducting properties ($T_c$ and shielding volume fraction) becomes the highest for $y = 0.4$, and superconductivity seems to be suppressed for further In doping. Therefore, the trend of In-doping effect on superconductivity in the $AgSnBiTe_3$ system is different from that in the SnTe and PbTe systems. We consider that the suppression of superconductivity is caused by the solution limit of In rather than the changes in electronic states.

We tried to synthesise samples with $y > 0.5$, but a high-purity sample was not obtained. In addition, the Bi amount for $y = 0.5$ deviates from the nominal value with a large error, which means that the sample with $y = 0.5$ also contains inhomogeneity larger than those in $y < 0.5$. Seeing lattice constants for In-doped samples, we find that the lattice constant is smaller than the end member of InTe with $a \sim 6.175$ Å[8]. Therefore, we consider that the suppression of superconductivity observed for $y = 0.5$ is due to the increase in inhomogeneity in the sample, and the solution limit of In for $AgSnBiTe_3$ is around $y = 0.4$ under pressures up to 2 GPa. As a fact, $y = 0.4$ samples show degradation of superconducting properties by passing time. Thus, in this study, investigations of superconducting properties of $y = 0.4$ have been performed in 24 hours after the high-pressure synthesis. Because the (Ag,In,Sn,Bi)Te system is a kind of disordered system with high configurational entropy of mixing ($\Delta S_{mix}$), we briefly discuss the possible explanation of the solution limit of In in (Ag,In,Sn,Bi)Te. As established in a field of high-entropy alloys, the phase of multiple-element system can be stabilized owing to high $\Delta S_{mix}$[29,30], which decreases Gibbs free energy, $\Delta G = \Delta H - T\Delta S$, at high temperatures, where $H$ is enthalpy. $\Delta S_{mix}$ values for $Sn_{1-2x}(AgBi)_xTe$ and $(AgSnBi)_{(1-y)/3}In_yTe$ were calculated using $\Delta S_{mix} = -R\Sigma^i c_i \ln c_i$, where $c_i$ and $R$ are the atomic fraction of component $i$ and the gas constant, respectively. As shown in Fig. 9a, for $Sn_{1-2x}(AgBi)_xTe$, $\Delta S_{mix}$ is relatively high with a wide range of $x$, which would be the reason why the phases could be obtained with a wide range of $x$ without the use of high-pressure synthesis. For In doping, we selected $AgSnBiTe_3$ ($x = 0.33$) as a parent phase in this study. For $(AgSnBi)_{(1-y)/3}In_yTe$, $\Delta S_{mix}$ becomes the highest at $y = 0.3$, and it decreases with further In substitution, as shown in Fig. 9b. Therefore, the possible explanation of the solution limit of In for $AgSnBiTe_3$ is as follows. Due to the small lattice constant of $\sim 6.2$ Å for $AgSnBiTe_3$, In substitution was challenging, but the phase was stabilized up to $y = 0.4$ by the use of high pressure and high $\Delta S_{mix}$. If we could get a phase with a higher In concentration, the superconducting phase diagram would be expanded from that shown in Fig. 7b.

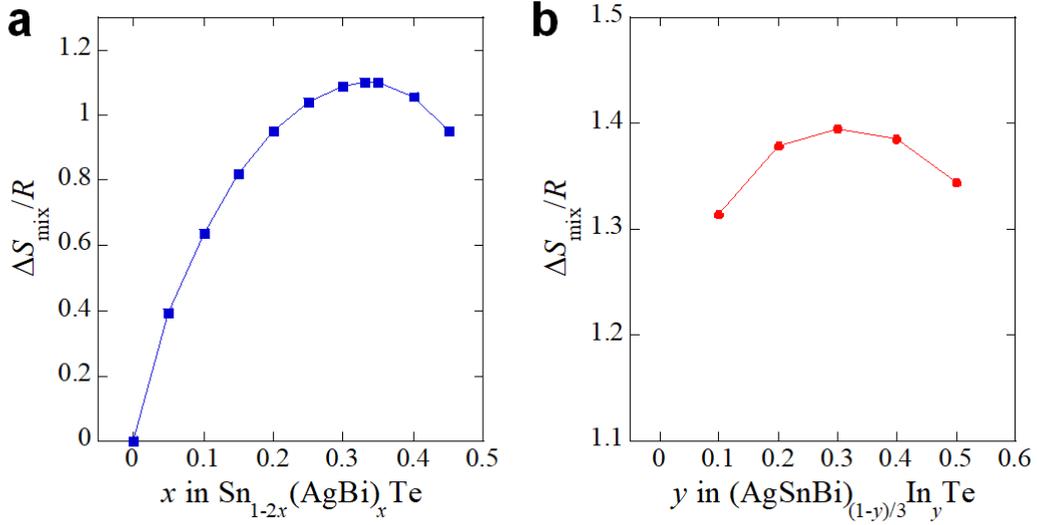

**Fig. 9. Configurational entropy of mixing ($\Delta S_{mix}$). a.** $x$ dependence of $\Delta S_{mix}$ for $Sn_{1-2x}(AgBi)_xTe$. **b.** $y$ dependence of $\Delta S_{mix}$ for $(AgSnBi)_{(1-y)/3}In_yTe$.

## Summary

We have synthesised polycrystalline samples of $Sn_{1-2x}(AgBi)_xTe$ and $(AgSnBi)_{(1-y)/3}In_yTe$ to explore a new candidate phase of topological superconductor. For $Sn_{1-2x}(AgBi)_xTe$, single-phase samples were obtained with a wide range of $x$. According to band calculations, we confirmed that the $Sn_{1-2x}(AgBi)_xTe$ system is basically possessing band inversion and topologically preserved electronic states. To investigate the effects of In substitution, we selected $x = 0.33$ ($AgSnBiTe_3$) as a parent phase. For $(AgSnBi)_{(1-y)/3}In_yTe$, In-doped samples were obtained for $y = 0$–$0.5$ by high-pressure synthesis, and superconductivity was observed for $y = 0.2$–$0.5$. For $y = 0.4$, specific heat investigation confirmed the emergence of bulk superconductivity. Although the current study is the material exploration with polycrystalline samples, we expect that the single crystals of (Ag,In,Sn,Bi)Te are grown in the next step, and characteristics including surface states, which are expected for a topological superconductor, are experimentally examined by surface-sensitive probes like angle-resolved photoemission spectroscopy.

## Methods

Polycrystalline samples of $Sn_{1-2x}(AgBi)_xTe$ ($x = 0.00, 0.05, 0.10, 0.15, 0.20, 0.25, 0.30, 0.33, 0.35, 0.40, 0.45$) and $(AgSnBi)_{(1-y)/3}In_yTe$ ($y = 0.0, 0.1, 0.2, 0.3, 0.4, 0.5$) were prepared by a melting method in an evacuated quartz tube. Powders or grains of Ag (99.9% up), Sn (99.99%), Bi (99.999%), In (99.99%), and Te (99.999%) were mixed and melted in an evacuated quartz tube at 900°C for 10 hours, followed by furnace cooling to room temperature. For $(AgSnBi)_{(1-y)/3}In_yTe$, the obtained samples were powdered and annealed in a high-pressure-synthesis instrument under 2 GPa at 500 °C for 30 minutes. A cubic-anvil-type 180-ton press was used, and the sample sealed in a BN crucible was heated by carbon heater.

The phase purity and the crystal structure of $Sn_{1-2x}(AgBi)_xTe$ and $(AgSnBi)_{(1-y)/3}In_yTe$ were examined by laboratory X-ray diffraction (XRD) by the $\theta$-$2\theta$ method with a Cu-K$_\alpha$ radiation on a MiniFlex600 (RIGAKU) diffractometer equipped with a high-resolution detector D/tex Ultra. The schematic images of crystal structures were drawn by VESTA[31] using a structural data refined by Rietveld refinement using RIETAN-FP[32]. The actual compositions of the examined samples were analysed using an energy dispersive X-ray spectroscopy (EDX) on TM-3030 (Hitachi).

The temperature dependence of magnetic susceptibility was measured using a superconducting quantum interference devise (SQUID) on MPMS-3 (Quantum Design) after zero-field cooling (ZFC) with an applied field of 10 Oe. The temperature dependence of electrical resistivity was measured by a four-probe method with an applied DC current of 1 mA on PPMS (Quantum Design) under magnetic fields. We used Ag paste and Au wires (25 μm in diameter) for the four-probe setup. The temperature dependence of specific heat was measured under 0 and 5 T by a relaxation method on PPMS. The resistivity and specific heat measurements were performed using a $^3$He probe system (Quantum Design). Hall coefficient was measured by four-probe setup on PPMS at low temperatures. Hall coefficient was estimated from the slope in the magnetic field dependence of Hall voltage. Seebeck coefficient at room temperature was measured under steady-state, where the thermo-electromotiveforce ($\Delta V$) and the temperature difference ($\Delta T$) were simultaneously measured., and the $S$ was determined from the slope of $\Delta V$ / $\Delta T$.

First principles band calculations were performed using WIEN2k package[33,34]. The electronic density of PbTe and SnTe was self-consistently calculated within the modified Becke–Johnson potential[35] using a 12 × 12 × 12 $k$-mesh and $RK_{max}$ = 9 with the spin–orbit coupling included.

## Acknowledgements
The authors thank O. Miura for experimental supports. This work was partly supported by JSPS KAKENHI (Grant No. 18KK0076, 21H00151, and 21K18834) and Advanced Research Program under the Human Resources Funds of Tokyo (Grant Number: H31-1).

## Author contributions
Y.M. and Y.G. led the project. T.M. and Y.M. synthesised the samples. T.M. and Y.M. characterised the samples using XRD and EDX. T.M. performed the magnetization measurements.

T.M., K.H., A.Y., R.K., R.H., T.D.M., Y.A., and Y.M. performed electrical resistivity and Hall measurements. T.M., M.R.K., A.Y., R.H., T.D.M., Y.A., and Y.M. performed specific heat measurements. M.R.K., T.K., and Y.M. performed Seebeck coefficient measurements. Theoretical calculations were carried out by H.U. This work was supervised in whole by Y.M., A.Y., and Y.G. The manuscript was written by H.U. and Y.M. with input from all coauthors.

## Competing Interests
The authors declare no competing interests.

# **Supplemental Information**

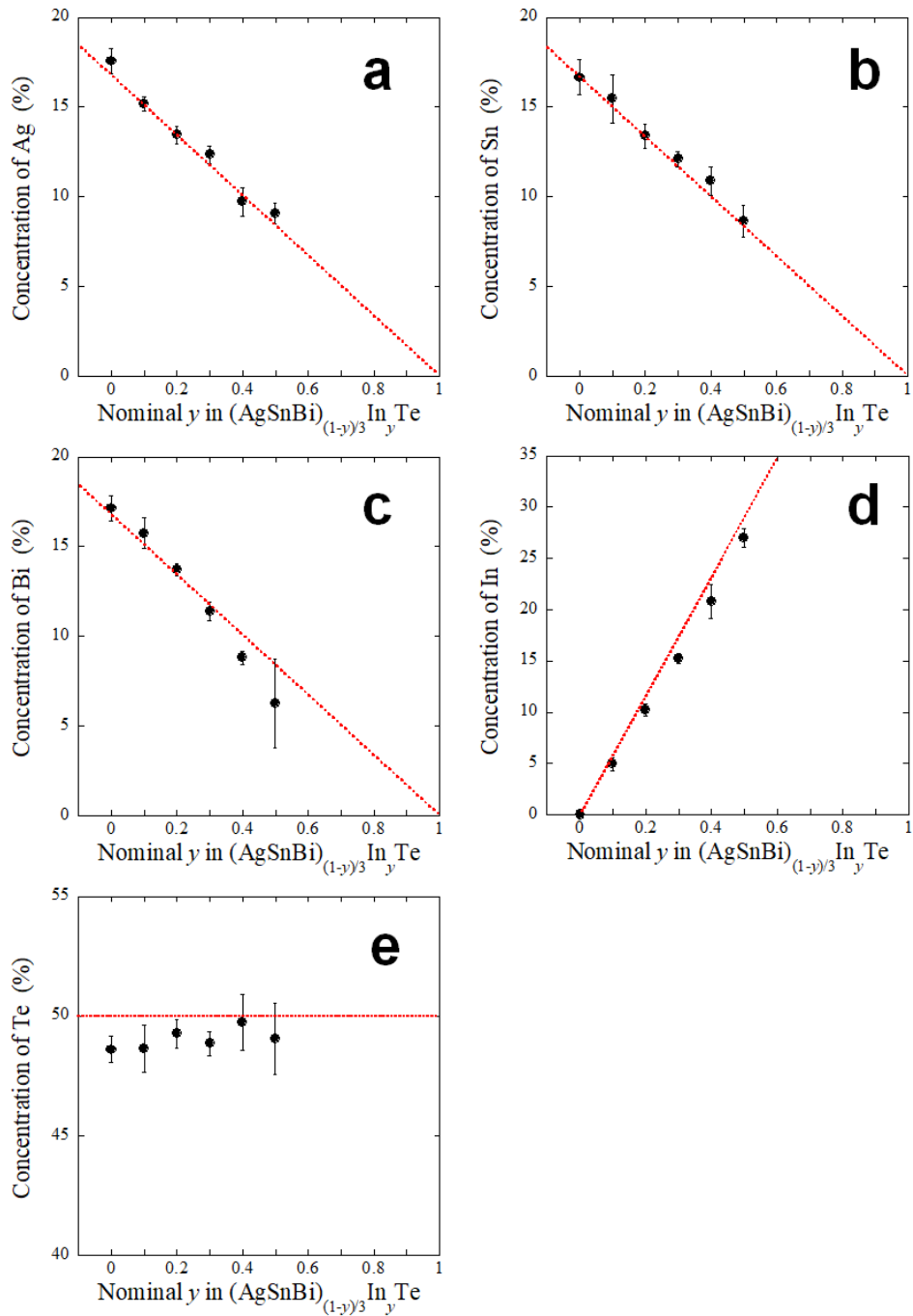

Fig. S1. EDX analysis results for $(AgSnBi)_{(1-y)/3}In_yTe$. Red dot lines indicate an ideal composition.

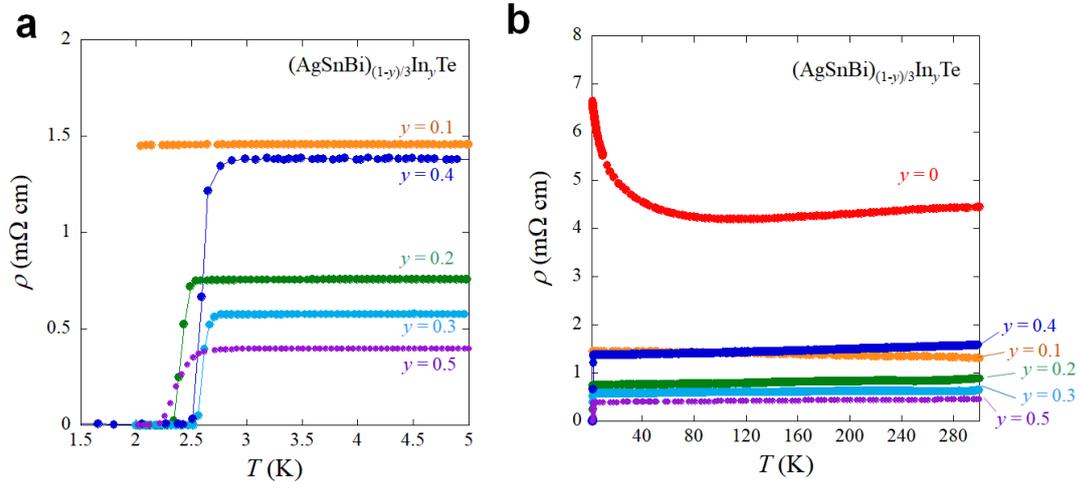

Fig. S2. Temperature dependences of electrical resistivity ($\rho$) for $(AgSnBi)_{(1-y)/3}In_yTe$.

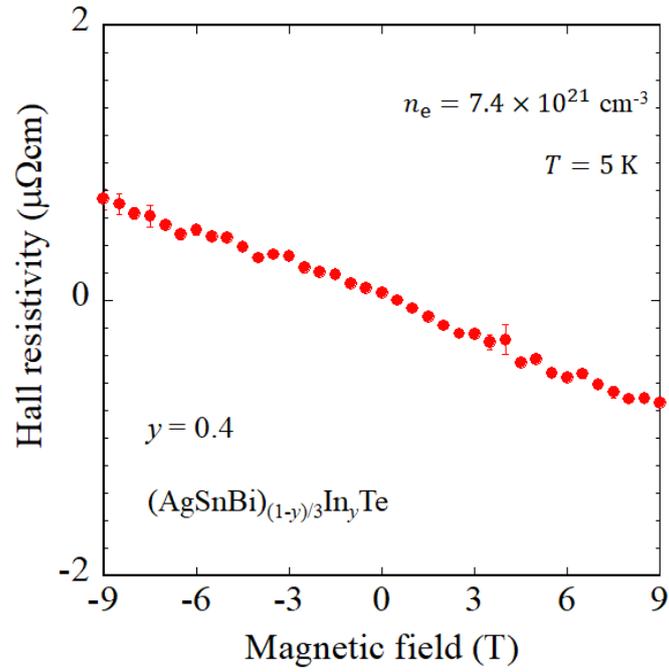

Fig. S3. Magnetic field dependences of Hall resistivity at 5 K for $y = 0.4$.